\documentclass[twocolumn,prb]{revtex4}
\usepackage{graphicx,psfrag}
\newcommand{\beq}{\begin{equation}}
\newcommand{\eeq}{\end{equation}}

\begin{document}
\title{Universal crossovers between entanglement entropy and thermal entropy}
\author{Brian Swingle}
\affiliation{Department of Physics, Harvard University, Cambridge, MA 02138}
\author{T. Senthil}
\affiliation{Department of Physics, Massachusetts Institute of Technology, Cambridge, MA 02139}
\begin{abstract}
We postulate the existence of universal crossover functions connecting the universal parts of the entanglement entropy to the low temperature thermal entropy in gapless quantum many-body systems.  These scaling functions encode the intuition that the same low energy degrees of freedom which control low temperature thermal physics are also responsible for the long range entanglement in the quantum ground state.  We demonstrate the correctness of the proposed scaling form and determine the scaling function for certain classes of gapless systems whose low energy physics is described by a conformal field theory. We also use our crossover formalism to argue that local systems which are ``natural" can violate the boundary law at most logarithmically.  In particular, we show that several non-Fermi liquid phases of matter have entanglement entropy that is at most of order $L^{d-1}\log{(L)} $ for a region of linear size $L$ thereby confirming various earlier suggestions in the literature. We also briefly apply our crossover formalism to the study of fluctuations in conserved quantities and discuss some subtleties that occur in systems that spontaneously break a continuous symmetry.
\end{abstract}
\maketitle
\section{Introduction}
Recently, an exchange of ideas between quantum information science and many-body physics has led to an improved understanding of the ``corner" of Hilbert space in which ground states of local Hamiltonians reside.  One of the most important tools for investigating the properties of many-body ground states is entanglement entropy, defined as the von Neumann entropy of the reduced density matrix of a spatial subsystem.  The ubiquitous presence of a boundary law for the entanglement entropy, as reviewed in Ref. \onlinecite{arealaw1,RevModPhys.80.517}, has provided a rough guide to the entanglement properties of quantum ground states.  This rough intuition led to a new class of quantum states generically called tensor network states Refs. \onlinecite{peps,terg,vidal_mera} as well as new insights into the classification and identification of many-body phases and phase transitions in Refs. \onlinecite{topent1,topent2,mps_classify1,mps_classify2,vidal_qcp}.

In this paper we show that by considering the relationship between thermal and entanglement entropy we can place significant constraints on ground state entanglement structure for ``natural" systems.  One of our main motivations is to characterize the possible violations of the boundary law for entanglement entropy at zero temperature.  There have been many constructions of anomalously entangled ground states in the quantum information community e.g. Ref. \onlinecite{chain_linear_ee}, but what do these have to do with ordinary quantum systems relevant for laboratory studies?  There are also motivations from the study of mutual information in quantum systems at finite temperature in Refs. \onlinecite{minfo,qmc_renyi1}. Interesting critical phenomena are visible in the mutual information, and a first step toward understanding these numerical results is a more complete understanding of the temperature dependence of the von Neumann entropy of a single region.

Our basic assumption that connects thermodynamics with entanglement is that the same low energy degrees of freedom are responsible for both long range entanglement and low temperature thermal physics.  To give a concrete example, in one dimensional relativistic critical systems, while the high energy physics contributes an area law term to the entanglement entropy, only the low energy modes contribute to the $\log{L}$ entanglement and low energy thermal properties as discussed in Refs. \cite{geo_ent,eeqft,korepin_thermee}.  Indeed, there is a universal crossover function that interpolates between the zero temperature entanglement entropy and the finite temperature thermal entropy of a given subregion (consisting of a single interval).  We study and generalize this crossover phenomenon in variety of critical systems in different dimensions.

In more detail, we will make the following assumptions throughout this paper.  We always study gapless systems since it is obvious (though not rigorous) that generic gapped phases obey a boundary law.  Our primary assumption is that there exists a universal crossover function that relates thermal and entanglement entropy.  This crossover function is only defined up to boundary law terms coming from high energy physics.  We also assume that the system does not possess extensive ground state degeneracy and that the Hamiltonian is not fine tuned (beyond the tuning necessary to reach criticality).  We will mostly consider translation invariant states, but we do discuss disordered states in Sec. VI.  In short, we want to consider sensible gapless ground states of local Hamiltonians, but in an effort to be precise, we give the above assumptions as sufficient criteria for ``sensibleness".  Finally, let us note that the renormalization group perspective on entanglement structure permeates our entire discussion.

We study the von Neumann entropy $S(L,T) = - \text{Tr}(\rho_L \log{\rho_L})$ of a real space region of linear size $L$ in $d$ spatial dimensions as a function of temperature, $L$, and region geometry.  Recall that at zero temperature most gapless quantum systems in $d > 1$ dimensions satisfy a boundary law for the entanglement entropy $S_L \sim L^{d-1}$ with the coefficient of this term non-universal (see Ref. \onlinecite{arealaw1}).  However, there are gapless systems that violate the boundary law for entanglement entropy including free fermions with a Fermi surface\cite{fermion1,fermion2,bgs_ferm1}, Landau Fermi liquids \cite{bgs_ferm2}, and Weyl fermions in a magnetic field at weak and strong coupling \cite{bgs_highe}.  These examples have an entanglement entropy $\sim L^{d-1} \log{(L)}$.

It is of enormous interest to generalize this result to understand the entanglement structure of {\em non-Fermi liquid} ground states of matter. Many such states share with the Fermi liquid the crucial feature that there are gapless excitations that reside at a surface in momentum space. However unlike in a Fermi liquid there is no description of these excitations in terms of a Landau quasiparticle picture.  Such states were suggested to also violate the area law for the entanglement entropy based on a heuristic argument that views that gapless momentum space surface as a collection of effective one dimensional systems \cite{bgs_ferm1}.   If the area law is indeed violated can the violation be stronger than in a Fermi liquid?

An example of a non-Fermi liquid state with a gapless surface in momentum space was studied numerically in Ref. \cite{qmc_renyi_critfs}.  The second Renyi entropy of a wavefunction (obtained by Gutzwiller projecting a free Fermi sea) for a gapless quantum spin liquid phase of an insulating spin system in two dimensions was calculated using Monte Carlo methods. The second Renyi entropy was shown to obey a behavior consistent with $L \log{(L)}$. Given the current limitations on system size it is hard to distinguish this from a power law violation of the area law. It is therefore important to have a general understanding of how seriously the area law can be violated in such a spin liquid state.

The quantum spin liquid phase discussed above is expected to be described by a low energy effective theory with a Fermi surface of emergent fermionic spin-$1/2$ particles (spinons) coupled to an emergent $U(1)$ gauge field. Similar effective field theories describe Bose metals, some quantum critical points in metals, and other exotic gapless systems.  In all these cases the violation of the boundary law is suggested by heuristic arguments. Based on the analogy with Fermi liquids we might guess that the violation is logarithmic\cite{bgs_ferm1}.  It is clearly important to have a firm argument for the correctness of this guess. Providing such an argument is one of the purposes of this paper.  What about other non-fermi liquid states where the effective theory is not yet understood? We will address a class of such states that have a critical Fermi surface with appropriate scaling properties \cite{critfs} to discuss the scaling constraints on their entanglement structure. In all these cases we argue that the $L^{d-1} \log{(L)}$ is the fastest possible parametric scaling with $L$ in $d$ dimensions.

Besides the von Neumann entropy, we also investigate the scaling behavior of fluctuations in conserved quantities as in Refs. \onlinecite{fermion1,bgs_renyi,numfluc_ent}.  Here the structure is slightly richer, but the basic conclusions are very similar.  In phases with unbroken symmetry, the fluctuations in the conserved quantity generating the symmetry scale no faster than $L^{d-1} \log{(L)}$ at zero temperature, again under the assumption that the same low energy modes responsible for thermal fluctuations also give rise to these zero temperature fluctuations.

This paper is organized as follows.  We begin with a discussion of the crossover behavior in the simplest conformally invariant case in $d$ dimensions.  Next we discuss the case of codimension one critical manifolds relevant for Fermi liquids, and then we discuss the general structure including higher codimension critical manifolds.  Finally, we turn to a discussion of fluctuations in conserved quantities.  We conclude with a discussion of possible violations of our scaling formalism.

\section{Scaling formalism: introduction}
\subsection{Conformal symmetry}
Consider a local quantum system with Hamiltonian $H$ at finite temperature $T = \beta^{-1}$ so that the entire system is in the mixed state $\rho(T) \propto \exp{(- \beta H)}$.  As $\beta \rightarrow 0$ we recover the ground state up to corrections exponential in the gap to the first excited state.  We will be exclusively interested in systems where $H$ is either in a gapless phase or at a critical point.  Thus we will always have some notion of scaling symmetry although we will often not have the full power of the conformal group.

Consider now a region $R$ of linear size $L$ inside a larger many-body system.  The complement of region $R$ is denoted $\bar{R}$.  The reduced density matrix of $R$ is
\beq
\rho_R(L,T) = \text{Tr}_{\bar{R}} (\rho(T))
\eeq
and the von Neumann entropy of this reduced density matrix is
\beq
S_R(L,T) = - \text{Tr}_R (\rho_R \log{\rho_R}).
\eeq
We will also be interested in generalizations of the von Neumann entropy called Renyi entropies labeled by a parameter $n$:
\beq
S_n = \frac{1}{1-n} \text{Tr}_R (\rho_R^n).
\eeq
The limit $n \rightarrow 1$ of $S_n$ is simply $S_R$, the von Neumann entropy of $\rho_R$.

Let us initially consider the special case of a conformal field theory in $d$ spatial dimensions.  Two simple limits exist.  As $L T \rightarrow 0$ (a non-universal velocity is suppressed) the von Neumann entropy recovers the usual entanglement entropy of the ground state.  We know from earlier studies that the entanglement entropy may contain a mixture of universal and cutoff dependent terms, see Refs. \onlinecite{ON_ent,free_ee_review,holo_ee} for representative calculations.  For example, the boundary law term, going as $L^{d-1}$ is non-universal, but there are universal logarithmic terms in $d=1,3,5,...$ dimensions.  In $d=2,4,...$ dimensions the universal logarithmic term is replaced by a constant term.  On the other hand, as $L T \rightarrow \infty$ we recover the usual thermal entropy going as $(L T)^d$.

Using our basic assumption we write the entropy of region $R$ as
\beq
S_R(L,T) = T^\phi f_R (LT) + ...
\eeq
where $...$ stands for the aforementioned addition or subtraction of non-universal terms involving the momentum cutoff $\Lambda$.  Let us now determine the properties of $f_R$ and the exponent $\phi$.  For the moment we suppress the region dependence writing $f_R$ as $f$. First, as $LT \rightarrow \infty$ we must recover the extensive thermal entropy and hence $f(x\rightarrow \infty) \sim x^d$.  This further implies that $\phi = 0$ to obtain the correct temperature dependence of the entropy.  In the opposite limit as $LT \rightarrow 0$ the only possibility for a non-zero and finite contribution is $f(x) \rightarrow \text{constant}$ or $f(x) \rightarrow \log{x}$.  The possibility of the logarithm is allowed since the $T$ appearing in the logarithm can be replaced by $\Lambda$ at the expense of a non-universal term.

We conclude that our scaling assumption is consistent with either a universal constant term or a universal logarithm in the entanglement entropy of a conformal field theory at zero temperature.  Indeed, both these possibilities are realized, the logarithmic term obtains for $d$ odd and the constant for $d$ even.  This is also an appropriate time to mention the possibility of shape dependence, for example, the fact that sharp corners produce logarithmic corrections is completely consistent with our scaling framework (see Refs. \onlinecite{corners_free_ee,free_ee_review}).  It is important to note that the constant term in odd dimensions is only meaningful in the absence of corners.  Other types of universal terms like $(\log{L})^p$ ($p\neq 1$) or $L^{d-1+\delta}$ ($\delta > 0$) are not allowed unless they violate our assumptions and are unrelated to the thermal physics.  Of course, this conclusion is very natural from the renormalization group point of view.

We briefly elaborate on this point and discuss the structure of high energy contributions in more detail.  Locality demands that all high energy contributions be proportional to integrals of local geometric data over the boundary.  Consider an entangling surface $\partial R$ in $d$ dimensions.  We may use coordinates $u^a$ ($a=1,...,d-1$) in terms of which the surface is $x^i(u^a)$ and the induced metric is $h_{ab} = \partial_a x^i \partial_b x^j \delta_{ij}$ (we only consider flat space here, the generalization is straightforward).  In addition to the intrinsic geometry of the surface we also have extrinsic geometry related to the embedding of the surface into flat space.  For example, a cylinder has extrinsic curvature but no intrinsic curvature where as a sphere has both.  The extrinsic geometry is controlled by the extrinsic curvature which is given in terms of the normal vector $n^i$ and the projector onto the surface $P^i_j = \delta^i_j - n^i n_j$ as $K_{ij} = P_i^k \partial_k n_j$.  Now an important constraint for global pure states is the requirement that $S(R) = S(\bar{R})$, and since we consider here only high energy contributions that are independent of the low energy physics, we may still demand this symmetry at finite temperature for the terms of interest.  Since the only difference between the boundary of $R$ and $\bar{R}$ is the direction of the normal $n$ we conclude that only even powers of $n$ can appear.  Thus only even powers of the extrinsic curvature and hence only even powers of derivatives can appear (the same is true for intrinsic terms).  Roughly speaking, we must form fully contracted invariants involving the normal vector and the gradient, but requiring the normal vector to appear with only even powers forces the same for gradients due to rotation invariance.  This explains the general even/odd structure of universal terms via a simple scaling argument e.g. Ref. \onlinecite{swingle_MI}.

Let $r$ denote the length scale of interesting along the RG flow.  The infinitesimal contribution to the entanglement entropy from degrees of freedom at scale $r$ is of the form described above:
\beq
r \frac{dS}{dr} = \frac{L^{d-1}}{r^{d-1}} \left( c_0 + c_2 \frac{r^2}{L^2} + ... \right)
\eeq
where we have used the appropriate logarithmic measure for $r$.  The presence of only even corrections comes from our argument above.  Performing this integral from the UV $r = \epsilon$ to the IR $r = L$ gives the desired structure.  More generally, one should cut this integral off at $\min{(L,1/T)}$ where, roughly speaking, the von Neumann entropy becomes thermal in nature.

Returning to our main development, the simplest example of such a crossover function occurs in $d=1$ conformal field theories where the single interval case is dictated by conformal invariance.  The result is
\begin{equation}
S(L,T) = \frac{c}{3} \log{\left(\frac{\beta \Lambda}{\pi}\sinh{\left(\frac{\pi L}{\beta}\right)} \right)},
\end{equation}
and we see immediately that this form is consistent with our general scaling hypothesis.  The high energy cutoff $\Lambda$ can be shifted by a boundary law respecting term, but the thermal physics and long range entanglement is independent of the precise choice of $\Lambda$ (again up to a non-universal dimensionful conversion factor).

\subsection{Conformal field theories in $d > 1$: Holographic calculation}
We have already mentioned one concrete example of such a crossover function in $1+1$ dimensional conformal field theory.  In general the computation of such a crossover function is a highly non-trivial task for interacting conformal field theories in spatial dimensions $d > 1$.   However, one set of examples where a computation is possible is provided by holography.  For an introduction to this set of ideas, see Ref. \onlinecite{mcgreevy_ads_review}.  For our purposes it suffices to mention three facts.  First, certain strongly interacting  conformal theories living in flat space are dual to theories of gravity in a curved higher dimensional space known as Anti-de Sitter (AdS) space .  The UV of the conformal field theory lives at the boundary of the gravitational spacetime.  Second, there is a simple prescription to compute the entropy in such theories, at least in a special limit.  Briefly, we must compute the ``area" (in Planck units) of a minimal ``surface" in the extended gravitational geometry that terminates in the UV on the boundary of the region of interest.  Third, finite temperature effects are dual to placing a black hole in the gravitational spacetime.

Putting all these facts together permits a geometric calculation of the entropy of the field theory that is precisely of the form we assumed.  We will now give the details of this calculation.  Consider first the case of CFT$_{1+1}$.  $x$ and $t$ are the CFT directions while $r$ is the emergent scale coordinate ($r\rightarrow 0$ is the UV boundary).  The gravitational geometry in this case is the AdS$_{2+1}$ black hole with metric
\begin{equation}
ds^2 = \frac{L_{\Lambda}^2}{r^2} \left( - f dt^2 + \frac{1}{f} dr^2 + dx^2 \right)
\end{equation}
where $f(r) = 1 - r^2/r_0^2$.  We study minimal curves $r(x)$ terminating in an interval of length $\ell$ at finite temperature.  The minimal length is given by
\begin{equation}
J = \int^{\ell/2}_{-\ell/2} \frac{dx}{r} \sqrt{1 + (dr/dx)^2 f^{-1}}.
\end{equation}
Minimizing this length with respect to $r$ gives an equation of motion which may immediately be integrated to yield a conserved quantity
\begin{equation}
\frac{1}{r \sqrt{1 + (dr/dx)^2 f^{-1}}} =  \frac{1}{r_m}
\end{equation}
where $r_m$ is the maximum depth achieved by the curve.  Solving this for $dr/dx$ gives
\begin{equation}
dr/dx = \sqrt{\left(\frac{r_m^2}{r^2} - 1\right)f} = \frac{\sqrt{(r_m^2-r^2) f}}{r}.
\end{equation}

We may now rewrite the length as
\begin{equation}
J = 2 \int_{\epsilon}^{r_m} \frac{dr \,r}{\sqrt{(r_m^2 - r^2) f}}\frac{r_m}{r^2}
\end{equation}
while the parameter $r_m$ is determined from
\begin{equation}
\ell = 2 \int_{\epsilon}^{r_m} \frac{dr\, r}{\sqrt{(r_m^2 - r^2) f}}.
\end{equation}
$\epsilon$ is a UV cutoff.

Changing variables to $w = r^2/r_m^2$ allows us to rewrite the integral for $\ell$ as
\begin{equation}
\ell = r_0 \int^1_0 dw \frac{1}{\sqrt{(1-w)(\alpha^2 - w)}}
\end{equation}
where we have safely put $\epsilon = 0$ and $\alpha = r_0/r_m \geq 1$.  Performing the integral we obtain
\begin{equation}
\ell/r_0 = \log{\left( \frac{\alpha^2-1}{(\alpha-1)^2}\right)} = \log{\left(\frac{\alpha+1}{\alpha-1}\right)}.
\end{equation}
We may now solve for $\alpha$ in terms of $\ell$
\begin{equation}
\alpha = \frac{1+e^{-\ell/r_0}}{1 - e^{-\ell/r_0}}.
\end{equation}

Returning now to the length $J$ we find
\begin{equation}
J = \frac{r_0}{r_m} \int_{\epsilon}^1 dw \frac{1}{\sqrt{(1-w)(\alpha^2 - w)}} \frac{1}{w}.
\end{equation}
Doing this integral gives
\begin{equation}
J = 2 \log{\left(\frac{2 r_0}{\epsilon}\sinh{\left(\frac{\ell}{2 r_0}\right)}\right)}.
\end{equation}
To compute the entropy we append the factor $\frac{L_{\Lambda}}{4 G_N}$ and use the relation $c = \frac{3 L_{\Lambda}}{2 G_N}$ to obtain
\begin{equation}
S = \frac{c}{3} \log{\left(\frac{2 r_0}{\epsilon}\sinh{\left(\frac{\ell}{2 r_0}\right)}\right)}.
\end{equation}

We may now determine $r_0$ in terms of the temperature. Zooming in near the horizon and writing $r = r_0 - \rho$ we have
\begin{equation}
ds^2 \sim \frac{L_{\Lambda}^2}{r_0^2} \left( - \frac{2 \rho}{r_0} dt^2 + \frac{r_0}{2 \rho} d\rho^2 \right).
\end{equation}
Changing variables to $u = 2 \sqrt{\rho}$ the near horizon metric is brought into the form
\begin{equation}
ds^2 \sim - \frac{u^2}{r^2_0} dt^2 + du^2.
\end{equation}
Demanding periodicity in imaginary time gives $\beta/r_0 = 2 \pi$ or $2 r_0 = 1/(\pi T)$.  Plugging this into our entropy formula reproduces the usual crossover function
\begin{equation}
S(\ell,T) = \frac{c}{3} \log{\left(\frac{1}{\pi T \epsilon}\sinh{\left(\pi T \ell \right)}\right)}.
\end{equation}

Let us now consider regions in higher dimensional conformal field theories.  For a $d+1$ dimensional CFT the relevant metric is the AdS$_{d+2}$ black hole
\begin{equation}
ds^2 = \frac{L_{\Lambda}^2}{r^2} \left( - f dt^2 + \frac{1}{f} dr^2 + dx_d^2 \right)
\end{equation}
where $f(r) = 1 - r^{d+1}/r_0^{d+1}$.

We consider strip-like regions in the field theory of cross-section $A$ and width $\ell$ ($A$ has units of length$^{d-1}$).  The minimal surface area is
\begin{equation}
\sigma = A \int_{-\ell/2}^{\ell/2} \frac{dx}{r^d}\sqrt{1 + (dr/dx)^2 f^{-1}}.
\end{equation}
As before, we may immediately integrate the equation of motion to yield the conserved quantity
\begin{equation}
\frac{1}{r^d \sqrt{1 + (dr/dx)^2 f^{-1}}} = \frac{1}{r_m^d}.
\end{equation}
$r_m$ has the same meaning as before.  Solving for $dr/dx$ we find
\begin{equation}
dr/dx = \frac{\sqrt{\left(r_m^{2d} - r^{2d}\right)f}}{r^d}.
\end{equation}

Putting these facts together gives two integrals
\begin{equation}
\sigma = 2 A \int_{\epsilon}^{r_m} \frac{dr\, r^{d}}{\sqrt{\left(r_m^{2d} - r^{2d}\right)f}}\frac{1}{r^{d}}\frac{r_m^{d}}{r^{d}} \label{stripA}
\end{equation}
and
\begin{equation}
\ell = 2 \int_{\epsilon}^{r_m} \frac{dr \,r^{d}}{\sqrt{\left(r_m^{2d} - r^{2d}\right)f}}. \label{stripL}
\end{equation}
We may set $\epsilon = 0$ in the integral for $\ell$ which gives a cutoff independent relation $r_m = r_m(r_0,\ell)$.  Of course, when $r_0 \rightarrow \infty (f \rightarrow 1)$ we have $r_m \propto \ell$.

For the area integral we write
\begin{equation}
\frac{\sigma}{2A} =\int_{\epsilon}^{r_m} dr \left[ \left(\frac{r_m^{d}}{r^d \sqrt{\left(r_m^{2d} - r^{2d}\right)f}} - \frac{1}{r^d} \right) + \frac{1}{r^d} \right]. \label{stripAren}
\end{equation}
We have essentially subtracted off the UV sensitive boundary law term so that the integral in parenthesis converges as $\epsilon \rightarrow 0$.
The $\epsilon$ dependence is now trivial to extract and we find
\begin{equation}
\frac{\sigma}{A} = \frac{2}{d-1} \frac{1}{\epsilon^{d-1}} + (\sigma/A)_{\mbox{fin}}(\ell,r_0).
\end{equation}
But this is of the required cross-over form: a universal cross-over function plus a boundary law respecting piece sensitive to UV physics.

\begin{widetext}
We may put this function into the precise form considered above by scaling the variables appropriately.  In (\ref{stripL}) set $r_m = \ell g( \ell T)$ and $w = r/r_m$ to obtain
\beq
1 = g(\ell T) \int^1_0 dw \frac{w^d}{\sqrt{1-w^{2d}}\sqrt{1-k(\ell T g(\ell T))^{d+1} w^{d+1}}}
\eeq
with $k$ some constant.  This is an implicit equation for the scaling function $g(x)$ that can easily be shown to have the properties claimed above.  In particular, it shows relativistic length-energy scaling.  Plugging this scaling form into (\ref{stripAren}) shows that $(\sigma/A)_{\text{fin}}$ has the form
\beq
T^{d-1} \frac{1}{(\ell T g(\ell T))^{d-1}} \left[2\int^1_0 dw \left(\frac{1}{w^d\sqrt{1-w^{2d}}\sqrt{1-k(\ell T g(\ell T))^{d+1} w^{d+1} }} - \frac{1}{w^d}\right) - \frac{2}{d-1} \right] = T^{d-1} f(\ell T).
\eeq
One comment is necessary, the overall factor $T^{d-1}$ differs from the $\phi$ obtained above simply because we are here working in a limit where $A$ is bigger than all other scales.  Repeating our general analysis above predicts $\phi=d-1$ since we must have $f(x\rightarrow \infty) \rightarrow x$.  Similarly, we have $f(x\rightarrow 0) \rightarrow x^{-(d-1)}$ to compensate the vanishing powers of $T$.
\end{widetext}

\subsection{Non-relativistic scale invariance}
In this section we discuss critical theories with dynamical exponent $z \neq 1$.  The dynamical exponent controls the relative scaling of space and time leading to the invariant form $L T^{1/z}$ where again we suppress a non-universal dimensionful parameter.  The thermal entropy of such a theory scales as $L^d T^{d/z}$ as follows simply from the requirement of dimensionlessness and extensivity.  Let us again introduce a universal scaling function following the assumptions above.  We write the entropy as $$ S(L,T) \sim T^{\phi} f(LT^{1/z}) + ...$$ and use the limit $LT^{1/z} \rightarrow \infty$ to establish that $f(x) \rightarrow x^d$ and $\phi = 0 $.  The rest of the analysis for the non-relativistic case is unchanged and again we are permitted at most universal constant or logarithmic terms in the entanglement entropy.

One can also perform a holographic computation in this setting using so called Lifshitz geometries.  In fact, the spatial part of the metric is unchanged at zero temperature, hence the structure of the entanglement entropy is identical.  For example, even space dimensions have subleading constants while odd space dimensions have subleading logs.  These zero temperature solutions may also be generalized into finite temperature black holes solutions, at least for certain values of $z$.  Ref. \onlinecite{koushik_lif_bh} contains a nice example of such a Lifshitz black hole with $d=z=2$ and metric
\begin{equation}
ds^2 = - f \frac{dt^2}{r^{2z}} + \frac{1}{f} \frac{dr^2}{r^2} + \frac{dx_2^2}{r^2}
\end{equation}
where $f = 1-r^2/r_0^2$.  As claimed, the only difference between this metric and the relativistic examples above is in the $r$ dependence of the $dt^2$ term and the different power of $r$ appearing in $f$.  The same manipulations establish a nearly identical crossover structure to the relativistic case except that the argument of all scaling functions is $\ell T^{1/2}$ instead of $\ell T$.  As always, a dimensionful constant has been suppressed.

\section{Scaling formalism: codimension $1$}
Now we turn to the case where the low energy degrees of freedom reside on a codimension $1$ subspace in momentum space.  By contrast, the scale invariant theories in the previous section had low energy degrees of freedom only a single point in momentum space (or finite set of isolated points).  Examples of systems with a codimension $1$ gapless surface includes Fermi liquids with a $d-1$ dimensional Fermi surface in $d$ dimensions, Bose metals, spinon Fermi surfaces, and much more.  Later we will consider the case of a general codimension gapless manifold.

\subsection{Review of Fermi liquids}
The low energy physics of a Fermi liquid is, for many purposes, effectively one dimensional (an exception to this rule is provided by zero sound which requires the full Fermi surface to participate).  Thus Fermi liquids violate the boundary law for entanglement entropy because one dimensional gapless systems violate the boundary law.  The anomalous term has been found to be universal in the sense that it depends only on the geometry of the interacting Fermi surface and not on any Landau parameters.  Remarkably, this term also controls the finite temperature entropy to leading order in $T/E_F$.  The universal part of the entanglement entropy and the low temperature thermal entropy are connected by a universal scaling function which can be calculated using one dimensional conformal field theory.

We work in $d = 2$ for concreteness.  Consider a real space region $A$ of linear size $L$ in a Fermi liquid with spherical Fermi sea $\Gamma$.  The entanglement entropy for this region scales as $S_L \sim k_F L \log{L} + L/\epsilon + ...$ with the boundary law violating term universal and the subleading term non-universal.  The precise value of the boundary law violating term is expressed in terms of the geometry of the real space boundary $\partial A$ and the Fermi surface $\partial \Gamma$ as
\begin{equation}
S_L = \frac{1}{2\pi} \frac{1}{12} \int_{\partial A} \int_{\partial \Gamma} dA_x dA_k |n_x \cdot n_k | \log{(L)}
\end{equation}
where $n_x$ and $n_k$ are unit normals to $\partial A$ and $\partial \Gamma$.  The intuition behind this formula is simply that the Fermi surface in a box of size $L$ is equivalent to roughly $k_F L$ gapless modes that each contribute $\log{L}$ to the entanglement entropy.  This formula is known as the Widom formula because of its relation to a conjecture of Widom in signal processing.  The Widom formula has not yet been rigorously proven, but it has been checked numerically and can be obtained simply from the one dimensional point of view.  To generalize to finite temperature we must replace the zero temperature one dimensional entanglement entropy by the general result at finite temperature given by
\begin{equation}
S_{1+1}(L,T) = \frac{c+\tilde{c}}{6} \log{\left( \frac{\beta v \Lambda}{\pi} \sinh{\left(\frac{\pi L}{\beta v}\right)}\right)}.
\end{equation}
Fermi liquids are described by many nearly free chiral fermions with $c=1$ and $\tilde{c} = 0$.  The marginal forward scattering interactions do not change the number of low energy modes, and hence the mode counting picture still works quantitatively.  However, we will only require a much more crude scaling assumption for our results.

\begin{widetext}
Following this one dimensional result a higher dimensional Fermi liquid also possesses a universal crossover between the low temperature thermal entropy and the universal part of the entanglement entropy.  This scaling function depends only on the geometry of the real space region $A$ and on the shape of the Fermi surface $\partial \Gamma$.  For a spherical Fermi surface and spherical real space region of radius $L$ this universal crossover function is given by
\begin{equation}
S(L,T) = \frac{1}{2\pi} \frac{1}{12} \int_{\partial A} \int_{\partial \Gamma} dA_x dA_k |n_x \cdot n_k | \log{\left( \sinh{\left(\frac{\pi 2 L |n_x \cdot n_k|}{\beta v_F(n_k)}\right)}\right)}.
\end{equation}
\end{widetext}

\subsection{Non-Fermi liquids}
We now consider the entanglement structure of non-Fermi liquid states.  We will restrict attention to the class of such states that have a codimension $1$ gapless surface in momentum space, a critical Fermi surface, but where there is no Landau quasiparticle description of the excitations.  A general scaling formalism has been developed for these states in Ref. \onlinecite{critfs}.  Of primary importance to us is the scaling of the thermal entropy.  Our considerations will apply to any non-Fermi liquid falling into the general scaling formalism of Ref. \onlinecite{critfs} irrespective of the detailed low energy theory.  However, to be concrete let us consider a Fermi surface coupled to a gauge field in $d=2$.

Recently there has been a controlled calculation of the properties of this system in terms of the gauge field dynamical critical exponent $z_b$ and the number $N$ of fermion flavors in Ref. \onlinecite{critfs_eps} following important earlier work in Refs. \onlinecite{critfs_problem,critfs_ising}.  The expansion parameters are $\epsilon = z_b - 2$ and $1/N$ with a controlled limit possible as $N \rightarrow \infty $ with $\epsilon N $ fixed.  This system was found to possess a critical Fermi surface.  Following the intuition for Fermi liquids, this system will violate the boundary law for entanglement entropy because of the presence of many gapless one dimensional degrees of freedom.  However, this situation is not a trivial generalization of the Fermi liquid case because the system lacks a quasiparticle description.

Thermodynamic quantities can be understood roughly in terms of many one dimensional gapless degrees of freedom on the Fermi surface with a dynamical critical exponent $z_f \neq 1$.  The thermodynamic entropy is predicted to be $S \sim k_F T^{1/z_f} $ ($k_F$ just measures the size of the Fermi surface).  Additionally, the low energy theory is such that only antipodal patches of the critical Fermi surface couple strongly to each other.  With our current knowledge, we cannot formulate the patch theory as a truly one dimensional theory, nevertheless thermodynamic quantities are correctly captured.  Furthermore, although the Fermi surface curvature must be kept in all existing formulations, this curvature enjoys a non-renormalization property which makes it into a kind of gauge variable: dispersing perpendicular to the patch normal is roughly like changing patches.  However, we reiterate that our results depend only on the thermodynamic entropy being given by $S \sim k_F T^{1/z_f} $.

As usual, we write the von Neumann entropy as
\begin{equation}
S(L,T) = k_F T^{\phi} f(L T^{1/z})
\end{equation}
where $z = z_f$ is the fermion dynamical critical exponent and $\phi$ is an exponent to be determined.  $k_F$ just measures the size of the Fermi surface.  It does not enter into the scaling argument in a non-trivial way.  Now from thermodynamics we know that for $x = LT^{1/z} \rightarrow \infty$ we must have $S \sim T^{1/z} L^2 $.  The $L$ dependence requires that $f(x) \sim x^2 $ as $x \rightarrow \infty$, and the $T$ dependence forces us to choose $\phi = -1/z$.  As $x = LT^{1/z} \rightarrow 0$ we must have $f(x)\sim x$ in order to cancel the diverging $T$ dependence.  More generally, we must have $f(x) = x \tilde{f}(x)$ with $\tilde{f}(x) = a + b \log{x}$ as before.  In particular, powers of log are not allowed because these would produce $T \rightarrow 0$ divergent terms that are supposed to be finite and universal.  This demonstrates that these non-Fermi liquid states may violate the boundary law at most logarithmically.

Let us also make some more detailed speculations.  The $z_b = 2$ critical Fermi surface actually corresponds to a marginal Fermi liquid, so for this theory with $N$ flavors we suspect that the boundary law violating term has the usual Fermi liquid form
\begin{equation}
S_L = N \times \frac{1}{2\pi} \frac{1}{12} \int_{\partial A} \int_{\partial \Gamma} dA_x dA_k |n_x \cdot n_k | \log{(L)}.
\end{equation}
At finite $\epsilon = z_b - 2$ we expect modifications of the prefactor due to $\epsilon$ dependent corrections.  However, it is likely that the geometric dependence of the integral remains unchanged.  Indeed, the different patches in the critical Fermi surface decouple much more strongly than they do in a Fermi liquid.  We also note this geometrical form has recently been verified in a holographic setup with log violations of the boundary law.   Depending on how precisely the critical Fermi surface is effectively one dimensional, a more structured crossover function of the form
\begin{equation}
S_L = \frac{1}{2\pi} \frac{T^{1/z} C(N,\epsilon)}{12} \int_{\partial A} \int_{\partial \Gamma} dA_x dA_k |n_x \cdot n_k | f_{\epsilon, N}(LT^{1/z})
\end{equation}
may be expected.  The function $f_{\epsilon,N}$ would play the role of $\log{(\sinh{(\pi x)})}$ in the Fermi liquid case.  This detailed geometric form may be too strong a requirement in general, but the general scaling form in the previous paragraph is certainly reasonable.

Similar scaling arguments can be made for the Renyi entropy $S_n = \frac{1}{1-n} \log{\left(\text{Tr}(\rho^n)\right)}$.  We know that the Renyi entropy at finite temperature has a specific relationship to the thermal entropy because of the simple scaling with $T$ of the finite temperature free energy.  The complete result is
\begin{equation}
S_n(T) = \frac{n - \frac{1}{n^{1/z_f}}}{n-1} \frac{1}{1+\frac{1}{z_f}} S(T)
\end{equation}
where $S(T)$ is the thermal entropy.  This $n$ dependence of the Renyi entropy actually holds for all $T$ and $L$ in the $z=1$ case of a Fermi liquid.  It would interesting to determine if this is also true for the $z\neq 1$ theory.  Because the Renyi entropy is potentially much easier to calculate numerically and analytically we believe it is a useful target for future work.

\section{Scaling formalism: general codimension}
As a simple example of what we have in mind, consider a free fermion system tuned so that the ``Fermi manifold" is of codimension $q$ in the $d$ dimensional momentum space.  $q=1$ is the generic case, a Fermi line in $d=2$ and a Fermi surface in $d=3$.  $q=2$ in $d=2$ and $q=3$ in $d=3$ correspond to Dirac points.  The interesting case of $q=2$ in $d=3$ is the problem of Fermi lines where the zero energy locus is one dimensional in the three dimensional momentum space.  The codimension is a useful parameter because it tells us the effective space dimension of the local excitations in momentum space.  For example, the Fermi surface case always $q=1$ indicating that the excitations are effectively moving in one dimension (radially).  Similarly, the case $q=2$ in $d=3$ corresponds to modes that move in two effective dimensions since there is no dispersion along the Fermi line.  Just as we could calculate the entropy of Fermi surfaces by integrating over the contributions of one dimensional degrees of freedom, we can obtain the entropy of these higher codimension systems by integrating over the contributions of dimension $q$ degrees of freedom.  We now make this explicit with a scaling argument.

Suppose that there exists a universal scaling function connecting the entanglement entropy to the thermal entropy for these codimension $q$ free Fermi systems.  If we assume a generic bandstructure then the dispersion may be linearized near the Fermi manifold yielding a dynamical exponent $z=1$.  Thus we write the von Neumann entropy of region $R$ as
\begin{equation}
S(L,T) = k^{d-q}_F T^{\phi} f(LT)
\end{equation}
where the factor of $k_F^{d-q}$ accounts for the size of the Fermi manifold.  The thermal entropy of such a system scales as $k_F^{d-q} L^d T^q$ and hence as usual we require $f(x) \rightarrow x^d$ as $x \rightarrow \infty$.  This also fixes $\phi = q - d < 0$.  Requiring regularity in the limit as $T \rightarrow 0$, we see that $f(x) \rightarrow x^{d-q} \tilde{f}(x)$ as $x \rightarrow 0$ where $\tilde{f}(x) = a + b \log{x}$.  Thus we discover that such systems may have a universal term proportional to $(k_F L)^{d-q}$ with either a constant or logarithmic prefactor.  We emphasize that this is precisely what one expects from integrating a $q$ dimensional contribution over the Fermi manifold.  In particular, we expect a constant prefactor for $q$ even and a logarithmic prefactor for $q$ odd because the $q$ dimensional system has $z=1$ and hence resembles a relativistic scale invariant theory of the type we considered in Sec. II.  These statements may be checked in the free fermion case because the entanglement entropy can be computed exactly, however, we defer a full discussion of this case to a future publication.

We conclude this section by noting that our conclusion is unmodified even if we have an interacting theory with a codimension $q$ gapless manifold and with general $z\neq 1$.  The scaling function has the form
\begin{equation}
S = k_F^{d-q} T^\phi f(LT^{1/z})
\end{equation}
with $\phi = (q-d)/z$ which is obtained by matching to the thermal entropy $k_F^{d-q} L^d T^{q/z}$.  Although we do not rigorously prove that the thermal entropy is always of this form, such a form does follow from a very general scaling analysis in momentum space and we know of no exceptions.  We still predict a universal term, constant or logarithmic, proportional to $(k_F L)^{d-q}$ at zero temperature.  For an example of such a transition, see Ref. \onlinecite{senthil_codim} which considered a quantum critical point between a line nodal metal and a paired superconductor.

\section{Fluctuations of conserved quantities}
In this section we give a brief description of our scaling formalism as applied to an interesting observable: the fluctuations of a conserved charge.  These considerations are motivated by the direct experimental accessibility of such fluctuations as well as by a desire to illustrate the general nature of our arguments.  As the primary example, consider a conserved number operator $N$ that may be written as a sum of local densities $N = \sum_r n_r$.  This operator commutes with the Hamiltonian and we may label energy eigenstates with different values of $N$.  Hence $N$ itself need have no fluctuations in the ground state.  However, we can consider the restricted operator $N_R = \sum_{r \in R} n_r$ which need not commute with the Hamiltonian and may have fluctuations.  An interesting measure of the correlations between $R$ and its environment is thus the quadratic fluctuations in $N_R$
\begin{equation}
\Delta N_R^2 = \langle (N_R - \langle N_R \rangle )^2 \rangle.
\end{equation}

We expect in a gapped phase of matter that these fluctuations satisfy a boundary law $\Delta N_R^2 \sim L^{d-1}$.  Gapless phases in higher dimensions also appear to satisfy a boundary law so long as the symmetry generated by $N$ is unbroken.  As usual, one dimension is an exception where it is known that $\Delta N_R^2 \sim \log{L}$ although the prefactor is not as universal as for the entanglement entropy i.e. it depends on the Luttinger parameter.  It has also been shown that Fermi liquids violate the boundary law for fluctuations in higher dimensions with $\Delta N_R^2 \sim \frac{1}{1+F_0}(k_F L)^{d-1} \log{(L)}$.

Because the discussion is so similar to the case of the entropy, we will not give any of the details here.  However, a few points are worth mentioning.  First, what we study is now the crossover between the zero temperature number fluctuations and the finite temperature number fluctuations as controlled by the thermodynamic compressibility.  Indeed, precisely such a crossover argument was used in Ref. \onlinecite{bgs_renyi} to argue that the number fluctuations in a Fermi liquid are modified by Fermi liquid parameters (unlike the entanglement entropy).  As before, we are allowed to subtract off any boundary law contribution since they can be generated by high energy degrees of freedom.  For relativistic scale invariant systems the finite temperature fluctuations go like $(LT)^d$ and we again find that only constant or logarithmic universal terms are allowed at $T=0$.  More generally, we conclude that $L^{d-1} \log{L}$ is the most severe violation of the boundary law that is possible without breaking a symmetry e.g. in critical Fermi surface systems that remain compressible (see Ref. \onlinecite{critfs}).  If the symmetry is broken then we expect $\Delta N_R^2 \sim L^d$ as follows from a mean-field wave function for a superfluid of the form $\prod_r e^{\alpha b_r^+}|\mbox{vac}\rangle$.

\begin{widetext}
For conformal field theories these statements may be straightforwardly checked using the fact that the conserved current $J^\mu$ has dimension $\Delta = d$.  To compute the charge fluctuations we must compute
\beq
\Delta N_R^2 = \int_R d^dx \int_R d^dy \langle J^0(x) J^0(y)\rangle = \int_R d^dx \int_R d^dy \frac{1}{|x-y|^{2d}} + \text{contact terms}.
\eeq
The contact terms are necessary to cancel a naive short distance volume scaling in the fluctuations.  Such a volume scaling from UV modes is unphysical because UV fluctuations are local and hence can only change the charge in region $R$ by appearing near the boundary.  Let us take $R$ to be a sphere of radius $L$.  We can do the integral over $y$ by changing variables to $w=y-x$ and integrating over $w$ in polar coordinates.  The limits of integration are from $r=|w|=0$ to $r = r_{\star}(\theta)=-|x| \cos{\theta} + \sqrt{L^2 - |x|^2 \sin^2{\theta}}$ where $\theta$ is the angle between $w$ and $x$.  Let us specialize to $d=2$ for simplicity.  The fluctuations now go like
\beq
 \int_R d^2x \int d\theta \frac{1}{r_{\star}(\theta)^2}
\eeq
where we have discarded the aforementioned UV divergent volume piece.  $r_{\star}$ remains non-zero for all $\theta$ unless $|x| \rightarrow L$, and it can be shown that the leading singularity is
\beq
\int d^2x \frac{1}{(L-|x|)^2} \sim \frac{L}{\epsilon}
\eeq
i.e. the boundary law.  If we further expand the $\theta$ integral in powers of $L-|x|$ we also find a sub-leading term giving rise to a logarithmic correction in agreement with our scaling result.
\end{widetext}

We wish to mention one final subtle point that arises in phases with broken symmetry and which is not properly captured in our thermodynamic treatment.  Thus consider a superfluid phase where the particle number symmetry is broken.  Besides the usual sounds modes that possess an energy scaling as $1/\ell$ ($\ell$ is system size), there is also a zero mode with energy levels scaling like $1/\ell^d$.  This zero mode plays an important role in finite size systems, like quantum Monte Carlo simulations, where it is responsible for insuring that although the symmetry is broken, the many-body state has a definite particle number.  In other words, it is related to the fact that the many-body ground state in a finite size system is a properly a cat state unless the symmetry is broken explicitly.  This zero mode is not easily visible in thermodynamics but it does affect the entanglement entropy and number fluctuations in an important way.

Let us ignore the sound modes and ask for a state that capture the dynamics of the zero mode.  Such a cat state has the form
\begin{equation}
| M \rangle \sim \int d \theta e^{- i M \theta} \bigotimes_r | \theta \rangle_r
\end{equation}
where $| \theta \rangle_r$ is a state of definite phase on site $r$.  We wish to trace out part of the system and compute the entropy of the remainder, but this problem has to be regulated because ambiguities are encountered in this procedure.  Consider a simpler system consisting $p$ states per site with a $Z_p$ symmetry relating them and where the many body state is of the form $\sum^{p}_{x = 1} \bigotimes_r | x \rangle_r$.  If we now trace out part of the system we find a reduced density matrix for region $R$ of the form
\begin{equation}
\rho_R = \mbox{tr}_{\bar{R}} \rho = \sum_x \bigotimes_{r \in R} | x \rangle \langle x |_r .
\end{equation}
Perfect correlation with the environment has rendered the reduced density matrix completely diagonal.  The entropy is now trivially $S = \log{p}$.  To connect this model to the superfluid, we need only estimate the effective value of $p$.  We do this by counting the effective number of orthogonal states in $R$.  Now the many-body coherent state of the form
\begin{equation}
| \theta \rangle = \bigotimes_{r \in R} e^{-|\alpha|^2/2} \sum_n \frac{ \alpha  e^{i n \theta}}{\sqrt{n!}} |n\rangle_r
\end{equation}
has an overlap with a neighboring state of the form
\begin{equation}
\langle \theta | \theta' \rangle = \exp{\left[|R| |\alpha|^2 \left(e^{-i(\theta-\theta')}-1\right)\right]}.
\end{equation}
Expanding in small $\theta-\theta'$, the first real term is $\exp{- |R| |\alpha|^2 (\theta-\theta')^2/2 }$ and hence states greater than $\Delta \theta \sim 1/\sqrt{|R| |\alpha|^2}$ are effectively orthogonal.  Hence we may take $p \sim 2 \pi/ \Delta \theta$ to give an entropy contribution of the form $\log{\sqrt{|R|}} = \frac{d}{2} \log{L}$.  We also see that this mean-field cat state captures the extensive in subsystem size number fluctuation while maintaining the ground state with definite particle number.

We can also pin the order parameter to remove the anomalous contribution to the entanglement. For the superfluid, there is now no anomalous entropy although there are still extensive number fluctuations. On the other hand, in an anti-ferromagnet we may pin the Neel field to point in a particular direction.  If we take as a mean field state
\begin{equation}
|\mbox{Neel} \rangle  = \bigotimes_{r \in A} |\uparrow \rangle_r \bigotimes | \downarrow \rangle_r
\end{equation}
then we see immediately that there is no anomalous entanglement and the fluctuations of $S_z$ are not extensive (they are zero).  However, the fluctuations of $S_x$ and $S_y$ are still extensive.  These simple considerations have been considerably developed in Ref. \onlinecite{PhysRevB.83.224410}.  A careful treatment including the interactions between the zero mode and the sound modes is also expected soon in Ref. \onlinecite{tower_sb_ee}.  They find that the coefficient of the logarithm is somewhat modified from the simple minded argument above.

\section{Potential violations}
Before concluding, let us point out some potential ways that one might violate the scaling forms we have developed.  A nice example is provided by a certain degenerate spin chain.  By fine tuning the strength of the nearest neighbor couplings as a function of position along the chain, Ref. \cite{chain_linear_ee} showed that it was possible to find a ground state with entanglement entropy that scaled as the length of the interval.  This was arranged by adjusting the couplings so that a real space RG procedure always coupled the boundary spin inside the region with a spin outside the boundary i.e. no spins formed singlets within the interval.  This required exponentially decaying couplings and is clearly not generic.

Another route is provided by large ground state degeneracy provided we use the completely mixed state.  Recently such systems have received a lot of attention due to results showing non-Fermi liquid behavior in certain holographic systems.  However, we emphasize that nothing is special about the holographic setting, it is but one example.  A simple condensed matter example with ``ground state degeneracy" is provided by the spin incoherent Luttinger liquid.  This system is a Luttinger liquid where the spin energy scale is much less than the charge energy scale.  At temperatures above the spin energy scale, the spin incoherent state emerges where the spin degrees of freedom are totally disordered.  Such a state has an extensive temperature independent entropy, but it also cannot be a true ground state.  To force the state to zero temperature, we must fine tune an infinite number of relevant operators (the entire spin Hamiltonian) to zero.  We call this an IR incomplete theory since it cannot be smoothly connected to zero temperature.  More generally, we can imagine intermediate scale RG fixed points that control the physics over a wide range of energies but which cannot be interpreted as ground states due to an infinite fine tuning.  We know one possibility is that such a state may have extensive entropy, but perhaps there are other possibilities where the entanglement entropy scales like $L^{d-1+\delta}$ for $0 < \delta <1$.

Another setting where violations might occur is in random systems.  In one dimension we know that even at infinite randomness fixed points the boundary law for the average entanglement entropy is violated is no worse than in the conformal case.  However, we do not know if infinite randomness fixed points would violate the boundary law in higher dimensions.  We expect any finite $z$ random fixed point will not, and we suspect that infinite randomness fixed points would not either, but we do not give a definite argument at this time.  We also note that there are considerable subtleties in these systems e.g. typical versus average values.  Since the entanglement entropy of a region has a probability distribution $p(S,L)$, it would be interesting to determine if the distribution was a function of $S/\log{(L)}$ only or something more complicated.  In any event, there are many open issues at such random fixed points, the thermodynamics does not obey the simple forms we have considered here, and so we do not have much else to say about these issues at this time.

Let us also briefly mention long range interactions.  If these interactions are due to massless fields with a non-singular action within the physical description e.g. fluctuating gauge fields or other critical bosonic modes, then a proper renormalization group description is possible and the entanglement entropy should have no additional anomalous structure.  Similarly, so long as the long range forces present in the system have such an interpretation, even if they must be introduced as auxillary fields, we might expect no new anomalies to appear.  On the other hand, consider the ``1d chain" where, in addition to nearest neighbor hoppings, every site can hop to every other site with the same strength.  Calling such a system one dimensional is a perversion, but it is an extreme form of long range interactions.  Clearly such a system can be expected to violate the one dimensional boundary law more than logarithmically.  The task that emerges is thus to understand where the crossover point is, as a function of the interaction range, to conventional one dimensional behavior. One quite interesting situation where these considerations are directly applicable is momentum space entanglement where the region $R$ is some subset of momentum space instead of position space. This topic, which has already received some preliminary attention, deserves its own exposition which we will present elsewhere.

Finally, we note that from the general codimension section our formalism can in some sense encompass states that violate the boundary law more seriously than logarithmically provided that $q$ is effectively less than $1$.  For example, $q=0$ roughly describes a state with gapless excitations everywhere in some region of finite measure in momentum space.  This is a lot like the situation with ground state degeneracy.  Nevertheless, as we have already said, we know of no example of a sensible ground state with $q < 1$.

\section{Discussion}
We have argued that entanglement entropy and thermal entropy may be connected via a universal crossover function in gapless phases and at critical points.  One major consequence of this assumption is that local quantum systems cannot violate the boundary law more than logarithmically.  However, we hasten to add that should our assumptions be violated, we have no objection.  In particular, possible loopholes escaping our conclusion include fine tuning in the Hamiltonian, systems with many degenerate ground states, and systems with long range interactions.  Models showing these characteristics may indeed be physically realistic in special cases, nevertheless, we argue that conventional gapless systems, even those with critical Fermi surfaces, will not violate the boundary law more than logarithmically.  Actually calculating the entanglement entropy in a model of a critical Fermi surface, perhaps in a $\epsilon$ expansion, and studying in more detail the entanglement properties for $q > 1$ are projects we leave for the future.

Many conventional quantum critical points that describe symmetry breaking transitions, like the $2+1$ XY critical point, fall into the category of conformal field theories discussed in Sec. II.  However, it was recently shown in Ref. \onlinecite{decon_qcp_ee} that ``deconfined" quantum critical points have a different entanglement structure due to proximate topologically ordered phases.  For example, Ref. \onlinecite{decon_qcp_ee} discussed the XY$^*$ critical point in $2+1$ dimensions which has the same correlation length exponent as the XY transition but a different anomalous dimension for the order parameter.  This arises because the order parameter $\Phi$ is actually composite $\Phi = b^2$ and it is the ``fractional" bosons $b$ that undergo the XY transition.  However, an important difference is that the $b$ bosons are coupled to a $Z_2$ gauge field and hence not all operators in the XY theory of the $b$s are gauge invariant.  Furthermore, there must always exist somewhere in the high energy spectrum gapped $Z_2$ vortices.  This is all to say that while the thermal entropy of XY and XY$^*$ are identical, they have different crossover functions and different entanglement properties at zero temperature.

A potentially profitable generalization of our work here would be to include in the scaling formalism the effect of relevant operators that move away from the critical point.  Similarly, it would be interesting to study the scaling structure of the full Renyi entropy in more detail.  Scale invariance fixes the high temperature Renyi dependence, but the zero temperature entanglement structure as a function of Renyi parameter could be rather rich.  Perhaps our scaling approach could shed some light on this structure.  The generalization to multiple regions is also open and is especially relevant to studies of mutual information.

It would also be quite interesting to develop a variational class of density matrices that encode the kind of crossover behavior we described here.  Since the mutual information always obeys a boundary law at finite temperature \cite{minfo}, such states could in principle look like a density matrix generalization of tensor network states, although presumably the bond dimension would have to grow as the temperature was lowered to account for systems that violate the boundary law at zero temperature.  Perhaps the new branching MERA approach\cite{b_mera} could help?  One could compute some of these crossover functions in field theory to see what sort of universal information is easily accessible.  We have already done the analogous calculations in holographic theories, but there it was already known that the entanglement entropy contains relatively little information due to the large $N$ limit.

\bibliography{cfs_ee}

\begin{thebibliography}{40}
\expandafter\ifx\csname natexlab\endcsname\relax\def\natexlab#1{#1}\fi
\expandafter\ifx\csname bibnamefont\endcsname\relax
  \def\bibnamefont#1{#1}\fi
\expandafter\ifx\csname bibfnamefont\endcsname\relax
  \def\bibfnamefont#1{#1}\fi
\expandafter\ifx\csname citenamefont\endcsname\relax
  \def\citenamefont#1{#1}\fi
\expandafter\ifx\csname url\endcsname\relax
  \def\url#1{\texttt{#1}}\fi
\expandafter\ifx\csname urlprefix\endcsname\relax\def\urlprefix{URL }\fi
\providecommand{\bibinfo}[2]{#2}
\providecommand{\eprint}[2][]{\url{#2}}

\bibitem[{\citenamefont{Eisert et~al.}(2010)\citenamefont{Eisert, Cramer, and
  Plenio}}]{arealaw1}
\bibinfo{author}{\bibfnamefont{J.}~\bibnamefont{Eisert}},
  \bibinfo{author}{\bibfnamefont{M.}~\bibnamefont{Cramer}}, \bibnamefont{and}
  \bibinfo{author}{\bibfnamefont{M.~B.} \bibnamefont{Plenio}},
  \bibinfo{journal}{Rev. Mod. Phys.} \textbf{\bibinfo{volume}{82}},
  \bibinfo{pages}{277} (\bibinfo{year}{2010}).

\bibitem[{\citenamefont{Amico et~al.}(2008)\citenamefont{Amico, Fazio,
  Osterloh, and Vedral}}]{RevModPhys.80.517}
\bibinfo{author}{\bibfnamefont{L.}~\bibnamefont{Amico}},
  \bibinfo{author}{\bibfnamefont{R.}~\bibnamefont{Fazio}},
  \bibinfo{author}{\bibfnamefont{A.}~\bibnamefont{Osterloh}}, \bibnamefont{and}
  \bibinfo{author}{\bibfnamefont{V.}~\bibnamefont{Vedral}},
  \bibinfo{journal}{Rev. Mod. Phys.} \textbf{\bibinfo{volume}{80}},
  \bibinfo{pages}{517} (\bibinfo{year}{2008}),
  \urlprefix\url{http://link.aps.org/doi/10.1103/RevModPhys.80.517}.

\bibitem[{\citenamefont{Verstraete et~al.}(2008)\citenamefont{Verstraete,
  Cirac, and Murg}}]{peps}
\bibinfo{author}{\bibfnamefont{F.}~\bibnamefont{Verstraete}},
  \bibinfo{author}{\bibfnamefont{J.}~\bibnamefont{Cirac}}, \bibnamefont{and}
  \bibinfo{author}{\bibfnamefont{V.}~\bibnamefont{Murg}},
  \bibinfo{journal}{Adv. Phys.} \textbf{\bibinfo{volume}{57}},
  \bibinfo{pages}{143} (\bibinfo{year}{2008}).

\bibitem[{\citenamefont{Gu et~al.}(2008)\citenamefont{Gu, Levin, and
  Wen}}]{terg}
\bibinfo{author}{\bibfnamefont{Z.-C.} \bibnamefont{Gu}},
  \bibinfo{author}{\bibfnamefont{M.}~\bibnamefont{Levin}}, \bibnamefont{and}
  \bibinfo{author}{\bibfnamefont{X.-G.} \bibnamefont{Wen}},
  \bibinfo{journal}{Phys. Rev. B} \textbf{\bibinfo{volume}{78}},
  \bibinfo{pages}{205116} (\bibinfo{year}{2008}).

\bibitem[{\citenamefont{Vidal}(2008)}]{vidal_mera}
\bibinfo{author}{\bibfnamefont{G.}~\bibnamefont{Vidal}},
  \bibinfo{journal}{Phys. Rev. Lett.} \textbf{\bibinfo{volume}{101}},
  \bibinfo{pages}{110501} (\bibinfo{year}{2008}).

\bibitem[{\citenamefont{Kitaev and Preskill}(2006)}]{topent1}
\bibinfo{author}{\bibfnamefont{A.}~\bibnamefont{Kitaev}} \bibnamefont{and}
  \bibinfo{author}{\bibfnamefont{J.}~\bibnamefont{Preskill}},
  \bibinfo{journal}{Phys. Rev. Lett.} \textbf{\bibinfo{volume}{96}},
  \bibinfo{pages}{110404} (\bibinfo{year}{2006}).

\bibitem[{\citenamefont{Levin and Wen}(2006)}]{topent2}
\bibinfo{author}{\bibfnamefont{M.}~\bibnamefont{Levin}} \bibnamefont{and}
  \bibinfo{author}{\bibfnamefont{X.-G.} \bibnamefont{Wen}},
  \bibinfo{journal}{Phys. Rev. Lett.} \textbf{\bibinfo{volume}{96}},
  \bibinfo{pages}{110405} (\bibinfo{year}{2006}).

\bibitem[{\citenamefont{Chen et~al.}(2011)\citenamefont{Chen, Gu, and
  Wen}}]{mps_classify1}
\bibinfo{author}{\bibfnamefont{X.}~\bibnamefont{Chen}},
  \bibinfo{author}{\bibfnamefont{Z.-C.} \bibnamefont{Gu}}, \bibnamefont{and}
  \bibinfo{author}{\bibfnamefont{X.-G.} \bibnamefont{Wen}},
  \bibinfo{journal}{Phys. Rev. B} \textbf{\bibinfo{volume}{83}},
  \bibinfo{pages}{035107} (\bibinfo{year}{2011}).

\bibitem[{\citenamefont{Fidkowski and Kitaev}(2011)}]{mps_classify2}
\bibinfo{author}{\bibfnamefont{L.}~\bibnamefont{Fidkowski}} \bibnamefont{and}
  \bibinfo{author}{\bibfnamefont{A.}~\bibnamefont{Kitaev}},
  \bibinfo{journal}{Phys. Rev. B} \textbf{\bibinfo{volume}{83}},
  \bibinfo{pages}{075103} (\bibinfo{year}{2011}).

\bibitem[{\citenamefont{Pfeifer et~al.}(2009)\citenamefont{Pfeifer, Evenbly,
  and Vidal}}]{vidal_qcp}
\bibinfo{author}{\bibfnamefont{R.~N.~C.} \bibnamefont{Pfeifer}},
  \bibinfo{author}{\bibfnamefont{G.}~\bibnamefont{Evenbly}}, \bibnamefont{and}
  \bibinfo{author}{\bibfnamefont{G.}~\bibnamefont{Vidal}},
  \bibinfo{journal}{Phys. Rev. A} \textbf{\bibinfo{volume}{79}},
  \bibinfo{pages}{040301} (\bibinfo{year}{2009}).

\bibitem[{\citenamefont{{Vitagliano} et~al.}(2010)\citenamefont{{Vitagliano},
  {Riera}, and {Latorre}}}]{chain_linear_ee}
\bibinfo{author}{\bibfnamefont{G.}~\bibnamefont{{Vitagliano}}},
  \bibinfo{author}{\bibfnamefont{A.}~\bibnamefont{{Riera}}}, \bibnamefont{and}
  \bibinfo{author}{\bibfnamefont{J.~I.} \bibnamefont{{Latorre}}},
  \bibinfo{journal}{New Journal of Physics} \textbf{\bibinfo{volume}{12}},
  \bibinfo{pages}{113049} (\bibinfo{year}{2010}), \eprint{1003.1292}.

\bibitem[{\citenamefont{Wolf et~al.}(2008)\citenamefont{Wolf, Verstraete,
  Hastings, and Cirac}}]{minfo}
\bibinfo{author}{\bibfnamefont{M.~M.} \bibnamefont{Wolf}},
  \bibinfo{author}{\bibfnamefont{F.}~\bibnamefont{Verstraete}},
  \bibinfo{author}{\bibfnamefont{M.~B.} \bibnamefont{Hastings}},
  \bibnamefont{and} \bibinfo{author}{\bibfnamefont{J.~I.} \bibnamefont{Cirac}},
  \bibinfo{journal}{Phys. Rev. Lett.} \textbf{\bibinfo{volume}{100}},
  \bibinfo{pages}{070502} (\bibinfo{year}{2008}).

\bibitem[{\citenamefont{Melko et~al.}(2010)\citenamefont{Melko, Kallin, and
  Hastings}}]{qmc_renyi1}
\bibinfo{author}{\bibfnamefont{R.~G.} \bibnamefont{Melko}},
  \bibinfo{author}{\bibfnamefont{A.~B.} \bibnamefont{Kallin}},
  \bibnamefont{and} \bibinfo{author}{\bibfnamefont{M.~B.}
  \bibnamefont{Hastings}}, \bibinfo{journal}{Phys. Rev. B}
  \textbf{\bibinfo{volume}{82}}, \bibinfo{pages}{100409}
  (\bibinfo{year}{2010}).

\bibitem[{\citenamefont{Holzhey et~al.}(1994)\citenamefont{Holzhey, Larsen, and
  Wilczek}}]{geo_ent}
\bibinfo{author}{\bibfnamefont{C.}~\bibnamefont{Holzhey}},
  \bibinfo{author}{\bibfnamefont{F.}~\bibnamefont{Larsen}}, \bibnamefont{and}
  \bibinfo{author}{\bibfnamefont{F.}~\bibnamefont{Wilczek}},
  \bibinfo{journal}{Nuc. Phys. B} \textbf{\bibinfo{volume}{424}},
  \bibinfo{pages}{443} (\bibinfo{year}{1994}).

\bibitem[{\citenamefont{Calabrese and Cardy}(2004)}]{eeqft}
\bibinfo{author}{\bibfnamefont{P.}~\bibnamefont{Calabrese}} \bibnamefont{and}
  \bibinfo{author}{\bibfnamefont{J.}~\bibnamefont{Cardy}}, \bibinfo{journal}{J.
  Stat. Mech.} \textbf{\bibinfo{volume}{04}}, \bibinfo{pages}{06002}
  (\bibinfo{year}{2004}).

\bibitem[{\citenamefont{Korepin}(2004)}]{korepin_thermee}
\bibinfo{author}{\bibfnamefont{V.~E.} \bibnamefont{Korepin}},
  \bibinfo{journal}{Phys. Rev. Lett.} \textbf{\bibinfo{volume}{92}},
  \bibinfo{pages}{096402} (\bibinfo{year}{2004}),
  \urlprefix\url{http://link.aps.org/doi/10.1103/PhysRevLett.92.096402}.

\bibitem[{\citenamefont{Wolf}(2006)}]{fermion1}
\bibinfo{author}{\bibfnamefont{M.}~\bibnamefont{Wolf}}, \bibinfo{journal}{Phys.
  Rev. Lett.} \textbf{\bibinfo{volume}{96}}, \bibinfo{pages}{010404}
  (\bibinfo{year}{2006}).

\bibitem[{\citenamefont{Gioev and Klich}(2006)}]{fermion2}
\bibinfo{author}{\bibfnamefont{D.}~\bibnamefont{Gioev}} \bibnamefont{and}
  \bibinfo{author}{\bibfnamefont{I.}~\bibnamefont{Klich}},
  \bibinfo{journal}{Phys. Rev. Lett.} \textbf{\bibinfo{volume}{96}},
  \bibinfo{pages}{100503} (\bibinfo{year}{2006}).

\bibitem[{\citenamefont{Swingle}(2010{\natexlab{a}})}]{bgs_ferm1}
\bibinfo{author}{\bibfnamefont{B.}~\bibnamefont{Swingle}},
  \bibinfo{journal}{Phys. Rev. Lett.} \textbf{\bibinfo{volume}{105}},
  \bibinfo{pages}{050502} (\bibinfo{year}{2010}{\natexlab{a}}).

\bibitem[{\citenamefont{Swingle}(2010{\natexlab{b}})}]{bgs_ferm2}
\bibinfo{author}{\bibfnamefont{B.}~\bibnamefont{Swingle}}
  (\bibinfo{year}{2010}{\natexlab{b}}), \eprint{arXiv:1002.4635}.

\bibitem[{\citenamefont{Swingle}(2010{\natexlab{c}})}]{bgs_highe}
\bibinfo{author}{\bibfnamefont{B.}~\bibnamefont{Swingle}}
  (\bibinfo{year}{2010}{\natexlab{c}}), \eprint{arXiv:1003.2434}.

\bibitem[{\citenamefont{Zhang et~al.}(2011)\citenamefont{Zhang, Grover, and
  Vishwanath}}]{qmc_renyi_critfs}
\bibinfo{author}{\bibfnamefont{Y.}~\bibnamefont{Zhang}},
  \bibinfo{author}{\bibfnamefont{T.}~\bibnamefont{Grover}}, \bibnamefont{and}
  \bibinfo{author}{\bibfnamefont{A.}~\bibnamefont{Vishwanath}},
  \bibinfo{journal}{Phys. Rev. Lett.} \textbf{\bibinfo{volume}{107}},
  \bibinfo{pages}{067202} (\bibinfo{year}{2011}),
  \urlprefix\url{http://link.aps.org/doi/10.1103/PhysRevLett.107.067202}.

\bibitem[{\citenamefont{Senthil}(2008)}]{critfs}
\bibinfo{author}{\bibfnamefont{T.}~\bibnamefont{Senthil}},
  \bibinfo{journal}{Phys. Rev. B} \textbf{\bibinfo{volume}{78}},
  \bibinfo{pages}{035103} (\bibinfo{year}{2008}),
  \urlprefix\url{http://link.aps.org/doi/10.1103/PhysRevB.78.035103}.

\bibitem[{\citenamefont{{Swingle}}(2010{\natexlab{a}})}]{bgs_renyi}
\bibinfo{author}{\bibfnamefont{B.}~\bibnamefont{{Swingle}}},
  \bibinfo{journal}{ArXiv e-prints}  (\bibinfo{year}{2010}{\natexlab{a}}),
  \eprint{1007.4825}.

\bibitem[{\citenamefont{{Song} et~al.}(2011)\citenamefont{{Song}, {Rachel},
  {Flindt}, {Klich}, {Laflorencie}, and {Le Hur}}}]{numfluc_ent}
\bibinfo{author}{\bibfnamefont{H.~F.} \bibnamefont{{Song}}},
  \bibinfo{author}{\bibfnamefont{S.}~\bibnamefont{{Rachel}}},
  \bibinfo{author}{\bibfnamefont{C.}~\bibnamefont{{Flindt}}},
  \bibinfo{author}{\bibfnamefont{I.}~\bibnamefont{{Klich}}},
  \bibinfo{author}{\bibfnamefont{N.}~\bibnamefont{{Laflorencie}}},
  \bibnamefont{and} \bibinfo{author}{\bibfnamefont{K.}~\bibnamefont{{Le Hur}}},
  \bibinfo{journal}{ArXiv e-prints}  (\bibinfo{year}{2011}),
  \eprint{1109.1001}.

\bibitem[{\citenamefont{Metlitski et~al.}(2009)\citenamefont{Metlitski,
  Fuertes, and Sachdev}}]{ON_ent}
\bibinfo{author}{\bibfnamefont{M.~A.} \bibnamefont{Metlitski}},
  \bibinfo{author}{\bibfnamefont{C.~A.} \bibnamefont{Fuertes}},
  \bibnamefont{and} \bibinfo{author}{\bibfnamefont{S.}~\bibnamefont{Sachdev}},
  \bibinfo{journal}{Phys. Rev. B} \textbf{\bibinfo{volume}{80}},
  \bibinfo{pages}{115122} (\bibinfo{year}{2009}),
  \urlprefix\url{http://link.aps.org/doi/10.1103/PhysRevB.80.115122}.

\bibitem[{\citenamefont{Casini and Huerta}(2009)}]{free_ee_review}
\bibinfo{author}{\bibfnamefont{H.}~\bibnamefont{Casini}} \bibnamefont{and}
  \bibinfo{author}{\bibfnamefont{M.}~\bibnamefont{Huerta}},
  \bibinfo{journal}{Journal of Physics A: Mathematical and Theoretical}
  \textbf{\bibinfo{volume}{42}}, \bibinfo{pages}{504007}
  (\bibinfo{year}{2009}).

\bibitem[{\citenamefont{Ryu and Takayanagi}(2006)}]{holo_ee}
\bibinfo{author}{\bibfnamefont{S.}~\bibnamefont{Ryu}} \bibnamefont{and}
  \bibinfo{author}{\bibfnamefont{T.}~\bibnamefont{Takayanagi}},
  \bibinfo{journal}{Phys. Rev. Lett.} \textbf{\bibinfo{volume}{96}},
  \bibinfo{pages}{181602} (\bibinfo{year}{2006}),
  \urlprefix\url{http://link.aps.org/doi/10.1103/PhysRevLett.96.181602}.

\bibitem[{\citenamefont{{Casini} and {Huerta}}(2007)}]{corners_free_ee}
\bibinfo{author}{\bibfnamefont{H.}~\bibnamefont{{Casini}}} \bibnamefont{and}
  \bibinfo{author}{\bibfnamefont{M.}~\bibnamefont{{Huerta}}},
  \bibinfo{journal}{Nuclear Physics B} \textbf{\bibinfo{volume}{764}},
  \bibinfo{pages}{183} (\bibinfo{year}{2007}), \eprint{arXiv:hep-th/0606256}.

\bibitem[{\citenamefont{{Swingle}}(2010{\natexlab{b}})}]{swingle_MI}
\bibinfo{author}{\bibfnamefont{B.}~\bibnamefont{{Swingle}}},
  \bibinfo{journal}{ArXiv e-prints}  (\bibinfo{year}{2010}{\natexlab{b}}),
  \eprint{1010.4038}.

\bibitem[{\citenamefont{{McGreevy}}(2009)}]{mcgreevy_ads_review}
\bibinfo{author}{\bibfnamefont{J.}~\bibnamefont{{McGreevy}}},
  \bibinfo{journal}{ArXiv e-prints}  (\bibinfo{year}{2009}),
  \eprint{0909.0518}.

\bibitem[{\citenamefont{{Balasubramanian} and
  {McGreevy}}(2009)}]{koushik_lif_bh}
\bibinfo{author}{\bibfnamefont{K.}~\bibnamefont{{Balasubramanian}}}
  \bibnamefont{and}
  \bibinfo{author}{\bibfnamefont{J.}~\bibnamefont{{McGreevy}}},
  \bibinfo{journal}{\prd} \textbf{\bibinfo{volume}{80}}, \bibinfo{eid}{104039}
  (\bibinfo{year}{2009}), \eprint{0909.0263}.

\bibitem[{\citenamefont{Mross et~al.}(2010)\citenamefont{Mross, McGreevy, Liu,
  and Senthil}}]{critfs_eps}
\bibinfo{author}{\bibfnamefont{D.~F.} \bibnamefont{Mross}},
  \bibinfo{author}{\bibfnamefont{J.}~\bibnamefont{McGreevy}},
  \bibinfo{author}{\bibfnamefont{H.}~\bibnamefont{Liu}}, \bibnamefont{and}
  \bibinfo{author}{\bibfnamefont{T.}~\bibnamefont{Senthil}},
  \bibinfo{journal}{Phys. Rev. B} \textbf{\bibinfo{volume}{82}},
  \bibinfo{pages}{045121} (\bibinfo{year}{2010}),
  \urlprefix\url{http://link.aps.org/doi/10.1103/PhysRevB.82.045121}.

\bibitem[{\citenamefont{Lee}(2009)}]{critfs_problem}
\bibinfo{author}{\bibfnamefont{S.-S.} \bibnamefont{Lee}},
  \bibinfo{journal}{Phys. Rev. B} \textbf{\bibinfo{volume}{80}},
  \bibinfo{pages}{165102} (\bibinfo{year}{2009}),
  \urlprefix\url{http://link.aps.org/doi/10.1103/PhysRevB.80.165102}.

\bibitem[{\citenamefont{Metlitski and Sachdev}(2010)}]{critfs_ising}
\bibinfo{author}{\bibfnamefont{M.~A.} \bibnamefont{Metlitski}}
  \bibnamefont{and} \bibinfo{author}{\bibfnamefont{S.}~\bibnamefont{Sachdev}},
  \bibinfo{journal}{Phys. Rev. B} \textbf{\bibinfo{volume}{82}},
  \bibinfo{pages}{075127} (\bibinfo{year}{2010}),
  \urlprefix\url{http://link.aps.org/doi/10.1103/PhysRevB.82.075127}.

\bibitem[{\citenamefont{Senthil and Shankar}(2009)}]{senthil_codim}
\bibinfo{author}{\bibfnamefont{T.}~\bibnamefont{Senthil}} \bibnamefont{and}
  \bibinfo{author}{\bibfnamefont{R.}~\bibnamefont{Shankar}},
  \bibinfo{journal}{Phys. Rev. Lett.} \textbf{\bibinfo{volume}{102}},
  \bibinfo{pages}{046406} (\bibinfo{year}{2009}),
  \urlprefix\url{http://link.aps.org/doi/10.1103/PhysRevLett.102.046406}.

\bibitem[{\citenamefont{Song et~al.}(2011)\citenamefont{Song, Laflorencie,
  Rachel, and Le~Hur}}]{PhysRevB.83.224410}
\bibinfo{author}{\bibfnamefont{H.~F.} \bibnamefont{Song}},
  \bibinfo{author}{\bibfnamefont{N.}~\bibnamefont{Laflorencie}},
  \bibinfo{author}{\bibfnamefont{S.}~\bibnamefont{Rachel}}, \bibnamefont{and}
  \bibinfo{author}{\bibfnamefont{K.}~\bibnamefont{Le~Hur}},
  \bibinfo{journal}{Phys. Rev. B} \textbf{\bibinfo{volume}{83}},
  \bibinfo{pages}{224410} (\bibinfo{year}{2011}),
  \urlprefix\url{http://link.aps.org/doi/10.1103/PhysRevB.83.224410}.

\bibitem[{\citenamefont{Metlitski et~al.}()\citenamefont{Metlitski, Grover, and
  Qi}}]{tower_sb_ee}
\bibinfo{author}{\bibfnamefont{M.}~\bibnamefont{Metlitski}},
  \bibinfo{author}{\bibfnamefont{T.}~\bibnamefont{Grover}}, \bibnamefont{and}
  \bibinfo{author}{\bibfnamefont{X.-L.} \bibnamefont{Qi}}, \bibinfo{journal}{to
  appear}  (????).

\bibitem[{\citenamefont{{Swingle} and {Senthil}}(2011)}]{decon_qcp_ee}
\bibinfo{author}{\bibfnamefont{B.}~\bibnamefont{{Swingle}}} \bibnamefont{and}
  \bibinfo{author}{\bibfnamefont{T.}~\bibnamefont{{Senthil}}},
  \bibinfo{journal}{ArXiv e-prints}  (\bibinfo{year}{2011}),
  \eprint{1109.3185}.

\bibitem[{\citenamefont{{Evenbly} and {Vidal}}(2011)}]{b_mera}
\bibinfo{author}{\bibfnamefont{G.}~\bibnamefont{{Evenbly}}} \bibnamefont{and}
  \bibinfo{author}{\bibfnamefont{G.}~\bibnamefont{{Vidal}}},
  \bibinfo{journal}{Journal of Statistical Physics}
  \textbf{\bibinfo{volume}{145}}, \bibinfo{pages}{891} (\bibinfo{year}{2011}),
  \eprint{1106.1082}.

\end{thebibliography}

\end{document}